\documentclass{sig-alternate-05-2015}

\usepackage{cite}

\begin{document}

\setcopyright{acmcopyright}
\doi{}
\isbn{}
\conferenceinfo{}{}
\acmPrice{}
\conferenceinfo{}{}
\title{A Local-Global LDA Model for Discovering Geographical Topics from Social Media}
\numberofauthors{3}
\author{
\alignauthor
Siwei Qiang\\
       \affaddr{Shanghai Jiao Tong University}\\
       \email{qiangsiwei@sjtu.edu.cn}
\alignauthor
Yongkun Wang\\
       \affaddr{Shanghai Jiao Tong University}\\
       \email{ykw@sjtu.edu.cn}
\alignauthor Yaohui Jin\\
       \affaddr{Shanghai Jiao Tong University}\\
       \email{jinyh@sjtu.edu.cn}
}

\maketitle

\begin{abstract}
Micro-blogging services can track users' geo-locations when users check-in their places or use geo-tagging which implicitly reveals locations. This ``geo tracking'' can help to find topics triggered by some events in certain regions. However, discovering such topics is very challenging because of the large amount of noisy messages (e.g. daily conversations). This paper proposes a method to model geographical topics, which can filter out irrelevant words by different weights in the local and global contexts. Our method is based on the Latent Dirichlet Allocation (LDA) model but each word is generated from either a local or a global topic distribution by its generation probabilities. We evaluated our model with data collected from Weibo, which is currently the most popular micro-blogging service for Chinese. The evaluation results demonstrate that our method outperforms other baseline methods in several metrics such as model perplexity, two kinds of entropies and KL-divergence of discovered topics.
\end{abstract}



\section{Introduction}
Micro-blogging services such as Twitter and Weibo are very popular for people to share their status with geo-locations. These data with geo-locations can be easily collected from the web. Such textual data with geo-coordinates or geo-tagged locations usually contain landmark information (e.g., scenic spots or famous restaurants) or information on local events (e.g., movies, vocal concerts, exhibitions or sports games), and hence, can provide us rich and interpretable semantics on different locations. It is also possible to infer inherent geographic variability of topics across various locations.

There are a lot of nice researches on addressing the questions of how the information is created and shared in different geographic locations and how the spatial and linguistic characteristics of people vary across regions. However, the challenge is that the messages with geo-locations are mixed with overwhelming noisy messages of daily chats or expressions of personal emotions, which have little or no relations to the location context. For example, in the city of Shanghai, the Bund (or \emph{Waitan}) is a famous waterfront and one of the most popular scenic spots for tourists. However, even in such a spot, Weibo are still full of daily conversations and greetings such as `Good night' or `Have a nice weekend', which has no local semantics. Therefore, it is very difficult to discover meaningful geo-location topics by existing methods such as inferring occurrences of words from local posts.

This paper proposes an effective method to handle noisy messages and model geographical topics of different locations. The proposed method is based on the Latent Dirichlet Allocation (LDA) \cite{blei2003latent} topic model. The intuitive idea is, for a specific location, the words used by users are different between a) daily conversations; and b) the description of the landmark or local events. The former is relatively consistent across different locations, and denoted as \textit{global} context, while the words in b) varies by sites and are denoted as \textit{local} context. Our method takes the local and global contexts into consideration and models each word to be generated from either its local or the global topic distribution by its estimated probabilities. The proposed strategy is able to distinguish locally featured words from noises and improve the quality of discovered topics. Our evaluation shows its effectiveness for information filtering and geographical topic discovery.

\section{Related Work}
Two lines of related research are briefly reviewed. Firstly, in order to support location-aware information retrieval or to compare topics across geographical locations, several researches studies have focused on discovering topics from geographical regions \cite{li2009probabilistic, yin2011geographical, hong2012discovering, yuan2013and, liu2015spatio}. For example, Yin et al. proposes and compares three ways of modeling geographical topics, including a location-driven model, a text-driven model, and a joint model called LGTA \cite{yin2011geographical}. Hong et al. takes the Markovian nature of users' locations into account and identifies topics based on location and language \cite{hong2012discovering}. However, they all focused on finding topically coherent locations only, and cannot cope with the difficulty of handling noisy messages.

The second line relevant to our research is local word detection. Based on the premise that local words should have concentrated spatial distributions around their location centers, Backstrom et al. proposes a spatial variation model for analyzing geographic distribution of terms in search engine query logs \cite{backstrom2008spatial}, and this method has been used by Cheng et al. to decide whether a word is local or not \cite{cheng2010you}. However, it is demonstrated by Wu et al. that this method can be erroneous since it assumes one peak density distribution while many local words can have multiple peaks \cite{wu2015semantic}. Our work is different from the existing works in that, word locality is not generated directly but evaluated by the generation probability, and thus a word (e.g. car) can be both non-\emph{local} for a majority of locations and also \emph{local} for a few particular locations (e.g. automobile 4S shops).

\section{Method}

\subsection{Local-Global LDA Model}
In our scenario, each geo-located document $d$ is explicitly tagged with a location $l$, and contains a set of words $w_{d}$. A geographical topic $z$ is a meaningful theme shared by similar locations, and each location is associated with a topic distribution $p(z|l)$. 

We formalize our model based on the following intuitions. Firstly, words close in space are likely to be clustered into the same geographical topic. Therefore, topics are generated from locations instead of individual documents. Secondly, locally featured words have a more compact geographical scope. For example, `bravo' is a word for a performer, so that it is more possible to be used at a theater, a concert or a stadium rather than other places. On the contrary, noisy words (e.g., happy, love, city) can have a much wider spatial range. However, some words could be local for certain locations although these words are commonly used at many places. In our method, the role (local or non-local) of a word is determined by its generation probabilities of its local and global semantic contexts. Therefore, our model is named as Local-Global LDA model (or LGLDA).

\begin{figure}
\centering
\includegraphics{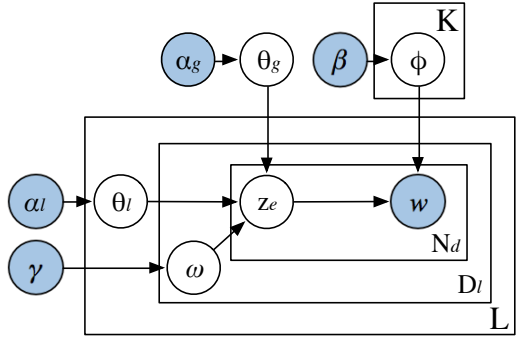}
\caption{Graphical representation of the proposed Local-Global LDA model.}
\end{figure}

The graphical representation of our model is shown in Figure 1. Shaded nodes indicate observed variables or priors, while light ones represent latent variables. In order to keep a small set of parameters for simplification, in our model there are one shared set of topics with two different distributions $\theta_{l}$ and $\theta_{g}$ for the local topics and the global topics, respectively. It might be interesting and reasonable to utilize two kinds of $\phi$ for words' local and global distributions corresponding to the two topic distributions, and we would like to study it in our future work. For a collection of $L$ locations, geo-tagged by $D$ documents, each contains $N$ words, the topic of each word can either be drawn from $\theta_{l}$ or from $\theta_{g}$. Topic assignment is denoted by $z_{e}$, and $e(=l/g)$ indicates whether it is drawn from the local or the global. Since micro-blog is length limited, it is likely to have focused concept. Therefore, if a document is location relevant, each word in it is more likely to be relevant. This relevance of each document is indicated by $\omega$, with binomial distribution prior $\gamma$. Finally, the word distribution in $K$ topics is denoted by $\phi$. $\alpha_{l}$, $\alpha_{g}$, $\beta$ are priors for $\theta_{l}$, $\theta_{g}$ and $\phi$, respectively.

In order to weight between local and global distributions, we add an additional parameter, named local-global weight ratio. Assume the local and global topic distributions are $\theta_{l}=[p_{l,1},...,p_{l,K}]$ and $\theta_{g}=[p_{g,1},...,p_{g,K}]$ respectively, topic assignment is drawn from a concatenated distribution $\theta$ in Equation (1).

\begin{equation}
\begin{aligned}
\theta &= \frac{\lambda}{\lambda+1}\theta_{l}p(e=l|w) \oplus  \frac{1}{\lambda+1}\theta_{g}p(e=g|w) \\
&=[\frac{\lambda p_{l,1}p_{w}^{(l)}}{\lambda+1},...,\frac{\lambda p_{w}^{(l)}}{\lambda+1},\frac{p_{g,1}p_{w}^{(g)}}{\lambda+1},...,\frac{p_{g,K}p_{w}^{(g)}}{\lambda+1}]
\end{aligned}
\end{equation}

When $\lambda$ is too large, the global word set is narrowed and ineffective for noise filtering,  while when $\lambda$ is too small, the size of local words is sparse and it fails to discover meaningful topics. Therefore, an appropriate $\lambda$ is crucial. In our experiment, it is optimized by estimating the model's perplexity (as illustrated in Figure 2 in Section 4.4). 

\subsection{Model Inference}
Like most Bayesian models, collapsed Gibbs sampling was used for model inference. We present the conditional probability of its latent variables $z_{e}, \theta_{l}, \theta_{g}, \phi$ and $w$ for sampling. Details are omitted for limited space. It is assumed that topic distributions $\theta_{l}$, $\theta_{g}$ and word distribution $\phi$ of each topic $k$ are drawn from dirichlet distributions of their respective priors $\alpha_{l}$, $\alpha_{g}$ and $\beta$, while locality relevance $\omega$ are drawn from a binomial distribution with prior $\gamma$. 

The conditional probability for sampling the topic assignment $z_{e}$ of each word is computed in Equation (2), where $z_{e,i} = k$ and $e_{i} = \kappa$ represent the assignments of the $i$th word to topic k, and mark the word as local if $\kappa=1$ otherwise non-local. $\lambda_{\kappa}=\frac{\lambda}{\lambda+1} (\kappa=l)$ or $\frac{1}{\lambda+1} (\kappa=g)$. $n_{-i,\kappa,k}$ represents the word count with locality assignment $\kappa$ and topic assignment $k$, and $-i$ means not including the $i$th word.

\begin{equation}
\begin{aligned}
p(z_{e,i} = k, e_{i} = \kappa | z_{-i}, e_{-i}) \propto \lambda_{\kappa} 
\cdot \frac{n^{(d_{i})}_{-i,\kappa,k}+\gamma_{\kappa}}{n^{(d_{i})}_{-i,\cdot,k}+\gamma_{l}+\gamma_{g}} \\
\cdot \frac{n^{(l_{i})}_{-i,\kappa,k}+\alpha_{\kappa}}{n^{(l_{i})}_{-i,\kappa,\cdot}+K \alpha_{\kappa}}
\cdot \frac{n^{(w_{i})}_{-i,\kappa,k}+\beta}{n^{(\cdot)}_{-i,\kappa,k}+W \beta},
\kappa \in \{l,g\}
\end{aligned}
\end{equation}

Consequently, the topic distribution of each location can be computed in Equation (3).

\begin{equation}
p(z_{i} = k|l_{i}) = \frac{n^{(l_{i})}_{1,k}+\alpha_{l}}{n^{(l_{i})}_{1,\cdot}+K \alpha_{l}}
\end{equation}

\section{Evaluation}

\subsection{Dataset}
The effectiveness of the proposed method was evaluated with the data (all in Chinese) collected from Weibo \cite{weibo}. Without loss of generality, we only focus on messages in Shanghai (the largest city in China and one of the largest cities in the world by population) in 2015. Since most of the locations have been tagged for only a few times, we only selected those locations with a considerable number of posts within a pre-defined spatial range. The original data was preprocessed by filtering out stop words, then nouns and verbs were extracted as valid words with a Chinese POS(part-of-speech) tagger. Messages with less than three valid words were eliminated. The details of this dataset are described in Table 1.

\begin{table}[h]
\centering
\caption{Dataset details}
\begin{tabular}{|l|c|} \hline
\textbf{Description} & \textbf{Value}\\ \hline
Total number of Weibos & 252173\\ \hline
Total number of locations & 1088\\ \hline
Average length of Weibos & 110.32\\
\hline\end{tabular}
\end{table}

\subsection{Comparison Methods}
We compare the proposed model LGLDA with the following other methods.

\begin{itemize}
\item \textbf{TF-IDF with K-means clustering (or TF-IDF).} \\
Weibos are firstly presented as tf-idf weighted vectors and aggregated by locations. Then, feature vector of each location is summed by documents and clustered by K-means. The center of each cluster represents a topic, and keywords are extracted by feature weight.
\item \textbf{LDA model with location aggregation (or LDA).} \\
Topic and word distributions are firstly calculated by standard LDA algorithm with all documents, without considering the geo-tagged locations. Then, the topic distribution of each location is calculated as the average over all documents geo-tagged with this location.
\item \textbf{Local LDA model (or LocalLDA).} \\
This method is similar to those in previous works but a simplified version in our scenario. Compared to LDA described above, the documents are calculated with geo-tagged locations in LocalLDA. In LocalLDA, there is only a local topic distribution, and other settings are kept the same with the LGLDA model.
\end{itemize} 

\subsection{Quantitative Measures}
In order to make a comparison between different methods, several quantitative measures are used. 

Perplexity is used to evaluate the performance of topic modeling. A lower perplexity score indicates better generalization performance of the model.
\begin{equation}
perplexity(D) = exp\{-\frac{\sum_{d \in D} \log p(w_{d})}{\sum_{d \in D} N_{d}}\}
\end{equation}
where $D$ is the test collection and $N_{d}$ is document length of document $d$.

Two entropies: the average topic entropy of each location and the average location entropy of each topic are used to measure the focusness of discovered topics. Each location should have a compact distribution on topics, while each topic should concentrate on a small set of locations.
\begin{equation}
entropy_{topic} = \sum_{L} (\sum_{K}p^{(l)}_{k}\log{p^{(l)}_{k}}) / L
\end{equation}
\begin{equation}
entropy_{location} = \sum_{K} (\sum_{L}p^{(k)}_{l}\log{p^{(k)}_{l}}) / K
\end{equation}
where $p^{(l)}_{k}$ and $p^{(k)}_{l}$ are the estimated probabilities of topic $k$ for location $l$ and location $l$ for topic $k$, respectively.

KL-divergence is used to measure the average distance of word distributions of all pairs of topics. The larger the average KL-divergence is, the more distinct the topics are.

\subsection{Settings and Results}

In our experiments, the number of topics was set at $K$ to 20 for all models (including K-means), $\alpha_{l}$ and $\alpha_{g}$ to 0.1, $\beta$ to 0.1, $\gamma_{l}$ and  $\gamma_{g}$ to 0.5 for all LDA-based models empirically and were run for 500 iterations. For LGLDA model, the local-global weight ratio was determined by model's perplexity as shown in Figure 2.  As can be seen, with the gradually increasing value of $\lambda$ from 0.1 to 20, the perplexity descended first and then ascended, and reached minimum at 0.6. Hence, 0.6 is the best value for $\lambda$, and was used in our experiments. Performance on the other two metrics also confirmed this selection.

\begin{figure}[b]
\centering
\includegraphics[width=3.6in]{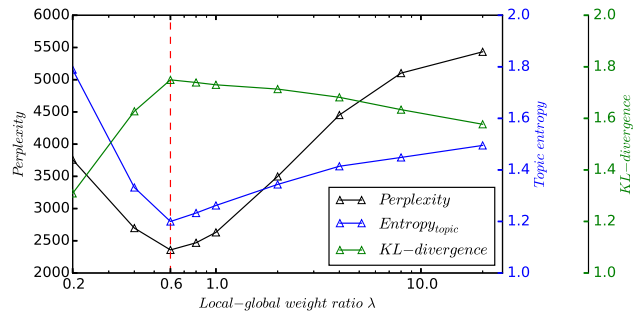}
\caption{The impact of the value of local-global weight ratio $\lambda$ on model's performance.}
\end{figure}

In order to validate the effectiveness of our model to distinguish between local words and noises, the locality score of Weibo defined as the ratio of the average generation probability of words from the local and global topic distribution was calculated in Equation (7). The higher the score is, the more relevant to local context the Weibos are. 
\begin{equation}
Locality(d) = \sum_{w \in d} p(z_{e,w}, e_{w}=l) / \sum_{w \in d} p(z_{e,w}, e_{w}=g)
\end{equation}

\begin{table*}[t]
\centering
\caption{Examples of Weibo text with locality score sampled from location tagged Waitan}
\begin{tabular}{|l|c|c|} \hline
\textbf{Weibo text} & \textbf{Locality score}\\ \hline
Cruising along Waitan while taking a close-up view of the Oriental Pearl Tower is a worthwhile trip. & 7.18\\ \hline
Breakfast at Xitang, lunch at Hangzhou, dinner at Waitan. An incredible day on the run. & 0.91\\ \hline
To deal with a difficult customer in the afternoon, I'd have to get up early to do data analysis. & 0.02\\
\hline\end{tabular}
\end{table*}

Three Weibos were sampled from the collection with distinguishable locality score with location tagged `Waitan' and shown in Table 2. The result is in accordance with expectation. Weibo with the highest score which includes several landmark names and location specific words is highly relevant, while the other two containing only few or no location featured words are weakly related or irrelevant. 

\begin{table*}[t]
\centering
\caption{Results of the comparative experiments}
\begin{tabular}{|l|c|c|c|c|} \hline
\textbf{Method} & \textbf{Perplexity} & \textbf{Topic entropy} & \textbf{Location entropy} & \textbf{KL-divergence}\\ \hline
LDA & 6904.11 & 2.9680 & 59.5100 & \textbf{2.2944}\\ \hline
LocalLDA & 5679.95 & 1.5156 & 31.2292 & 1.5694\\ \hline
LGLDA &  \textbf{2357.94} & \textbf{1.1998} & \textbf{24.2575} & \textbf{1.7494}\\
\hline\end{tabular}
\end{table*}

Table 3 gives the comparative results with other methods. Our model outperforms the others in perplexity, topic and location entropies, and KL-divergence. It achieves much lower perplexity since it can separate local and non-local words, thereby the documents are better organized. With noises filtered out, the discovered topics are more concentrated and representative for local semantics. However, the reason that KL-divergence of our model is lower than the LDA baseline is due to the location relevance constraint.

\newcommand{\tabincell}[2]{\begin{tabular}{@{}#1@{}}#2\end{tabular}}
\begin{table}[h]
\centering
\caption{A comparison of TOP3 topics discovered}
\begin{tabular}{|c|c|c|c|c|} \hline
\textbf{TF-IDF} & \textbf{LDA} & \textbf{LocalLDA} & \textbf{LGLDA}\\ \hline
\tabincell{c}{work,\\company,\\mood,\\women,\\teacher}  &
\tabincell{c}{love,\\feeling,\\mate,\\inside,\\teacher} & 
\tabincell{c}{work,\\mood,\\love,\\children,\\phone} & 
\tabincell{c}{work,\\company,\\mood,\\phone,\\city}\\ \hline
\tabincell{c}{university,\\teacher,\\effort,\\mood,\\paper}  &
\tabincell{c}{teacher,\\English,\\school,\\exam,\\culture} & 
\tabincell{c}{university,\\teacher,\\library,\\school,\\birthday} & 
\tabincell{c}{university,\\school,\\library,\\teacher,\\student}\\ \hline
\tabincell{c}{Waitan,\\city,\\night,\\Oriental\\ Pearl\\ Tower,\\restaurant}  &
\tabincell{c}{city,\\Waitan,\\Shanghai,\\Oriental\\ Pearl\\ Tower,\\international} & 
\tabincell{c}{Waitan,\\hotel,\\center,\\Oriental\\ Pearl\\ Tower,\\financial} & 
\tabincell{c}{Waitan,\\Oriental Pearl\\ Tower,\\Chenghuang\\ Temple,\\Nanjing Road,\\Huangpu River}\\
\hline\end{tabular}
\end{table}

Table 4 represents the comparison of the top3 topics discovered. The first topic (2nd row) is composed of broader words which can be viewed as noises, while the other two (3rd and 4th row) contain the semantics of education and tourist attractions. As can be seen, keywords discovered by our LGLDA model achieve the best relevance while keywords by other methods include more or less noises. From our results, Topic1 together with Topic4 accounts for 98.8\% of global topic distribution in LGLDA. With noises filtered out, locations are covered by less topics. For example, for Oriental Pearl Tower, the weight of Topic3 in LGLDA is 0.933 compared with a mixed topic constitution discovered by LocalLDA (0.425 for Topic1, 0.414 for Topic3 and 0.161 for others).

\section{Conclusion}
This paper proposes a method to combine local word filtering and geographical topic modeling into one framework. The proposed LDA-based model can effectively distinguish between local words and a variety of noisy daily interests by properly choosing the local-global weight ratio parameter. Results on Weibo collection show the effectiveness of our method over other baselines. 


\section{Acknowledgments}
This work was supported by the National Natural Science Foundation of China (Grant No. 61371084 and 91324202). Data and code will be published on Github after this review.

\bibliographystyle{ieeetr}
\bibliography{sigproc}

\end{document}